\documentclass[aps,prl,fleqn,twocolumn,superscriptaddress,floatfix]{revtex4}

\usepackage{graphicx}
\usepackage{amsmath,amssymb,amsfonts}

\begin{document}

\title{Transverse Velocity Dependence of the Proton-Antiproton Ratio 
         as a Signature of the QCD Critical Point}

\author{M.~Asakawa}
\affiliation{Department of Physics, Osaka University, Toyonaka 560-0043, Japan}

\author{S.~A.~Bass}
\affiliation{Department of Physics, Duke University, Durham, NC 27708, USA}

\author{B.~M\"uller}
\affiliation{Department of Physics, Duke University, Durham, NC 27708, USA}

\author{C.~Nonaka}
\affiliation{Department of Physics, Nagoya University, Nagoya 464-8602, Japan}

\date{\today}

\begin{abstract}
The presence of a critical point in the QCD phase diagram can deform the trajectories
describing the evolution of the expanding fireball in the $\mu_B-T$ phase diagram. If the
average emission time of hadrons is a function of transverse velocity, as microscopic
simulations of the hadronic freeze-out dynamics suggest, the deformation of the hydrodynamic
trajectories will change the transverse velocity ($\beta_{\rm T}$) dependence of the proton-antiproton
ratio when the fireball passes in the vicinity of the critical point. An unusual 
$\beta_{\rm T}$-dependence of the $\bar{p}/p$ ratio in a narrow beam energy window would thus 
signal the presence of the critical point.
\end{abstract}

\maketitle

Lattice-QCD simulations have shown that the transition between
the hadronic and quark-gluon plasma phases of quantum chromodynamics (QCD) at vanishing
baryon chemical potential $\mu_B$ is a crossover transition \cite{Aoki:2006we}. This raises the
question whether the crossover transition becomes a first-order phase transition for larger values 
of $\mu_B$. Several attempts have been made to locate the critical point, i.~e.\ the endpoint of the 
first-order transition line, in lattice simulations \cite{Fodor:2002km,Fodor:2004nz,Gavai:2004sd},
but its existence is still in doubt \cite{deForcrand:2007rq}. The inconclusive theoretical results have 
motivated plans for a systematic exploration of the properties of hot QCD matter as a function of the 
net baryon density by means of a collision energy scan at the Relativistic Heavy Ion Collider (RHIC) 
\cite{Stephans:2006tg,Lacey:2006bc}. The search for the QCD critical point also forms part of the motivation for the 
NA61 experiment \cite{Laszlo:2007ib}  at the CERN-SPS and for a new facility dedicated to the study 
of compressed baryonic matter at the Facility for Antiproton and Ion Research (FAIR) in Germany.  

Ideas for experimental signatures for the presence of the critical point have mostly focused 
on fluctuations in certain observables related to the order parameter of the chiral transition 
\cite{Stephanov:1998dy,Stephanov:1999zu}. General arguments lead one to believe that
such fluctuations are enhanced in the vicinity of the critical point. Unfortunately, several reasons 
throw doubt on the usefulness of fluctuation observables as practical signatures of the QCD critical
point. First, fluctuations are suppressed, compared to the static case, when the matter passes rapidly 
through the critical region during the expansion due to critical slowing down \cite{Berdnikov:1999ph}. 
Secondly, the hot matter does not freeze out at the critical point, but at a much lower temperature, when 
the critical fluctuations may well have been washed out. Finally, it is unclear in which observable 
fluctuations are most promising experimentally. In exploratory experiments at the CERN-SPS,
only fluctuations in the $K/\pi$ ratio at beam energies below 40 GeV/$A$ have shown signs 
of an unusual behavior \cite{Roland:2005pr,Roland:2006zz}.

Here we propose a possible signature of the presence of a critical point in the QCD phase
diagram, which may be more robust than fluctuations associated with the order parameter of
the chiral phase transition. Our idea is based on the observation that the critical point serves as 
an attractor of the hydrodynamical trajectories in the $\mu_B-T$ plane describing the expansion
of the hot matter \cite{Nonaka:2004pg}. We describe below how this focusing effect manifests
itself in an experimental observable.

The universality argument tells us that the critical exponents around second order phase 
transitions are determined only by the dimensionality and symmetry of the system. The QCD
critial point, if it exists, belongs to the same universality class as the 3-dimensional Ising 
model and liquid-gas phase transition \cite{Stephanov:1999zu}. The singular part of the 
thermodynamic variables near the critical point is a function of two variables, which can be 
mapped onto the variables characterizing the phase diagram of the 3-dimensional Ising model: 
the reduced temperature $r=(T-T_c)/T_c$ and the external magnetic field $h$. In the QCD 
phase diagram, the axis corresponding to $r$ points in the direction of the phase boundary; the 
direction of the axis associated with the variable $h$ is not known \cite{Berdnikov:1999ph,Hatta:2002sj}. 
However, it is clear that the critical region is more elongated along the $r$-direction, because the 
critical exponent associated with $r$ is larger than that associated with $h$ \cite{Berdnikov:1999ph}. 

The focusing effect can now be understood as follows.
The entropy density $s$ and the baryon density $n_b$ depend in different ways on $r$ and $h$. 
As a result, the ratio $s/n_b$, which is constant along an isentropic trajectory, assumes many 
different values in the vicinity of the critical point. Therefore, hydrodynamic trajectories for a 
range of different values of $s/n_b$ pass near the critical point, thus causing the focusing effect.

\begin{figure}[htb]
\centerline{\includegraphics[width=0.95\linewidth]{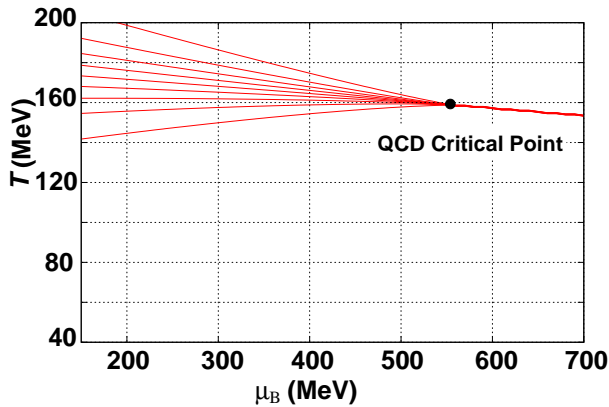}}
\caption{Critical region in the $\mu_B-T$ plane. The thick solid line to the right of the black dot 
shows the location of the phase boundary, starting at the critical point  $(\mu_c, T_c) = 
(550, 159)$ MeV. The thin lines to the left indicate the contours of equal values of the cross-over 
parameter $\tanh S_c$ between $-0.8$ and $+0.8$ in increments of 0.2.}
\label{fig1}
\end{figure}

The extent of the focusing region depends on the size of the critical region in the $\mu_B-T$ 
plane, in which thermodynamic susceptibilities are significantly enhanced by the critical exponents. 
The size of the attractive basin can, in principle, be determined by lattice-QCD simulations. At the 
moment this information is not available, as the location and even the existence of the critical point 
in QCD are not established. Model studies in simplified theories suggest that the size of the attractive 
region is sensitive to calculational details \cite{Schaefer:2006ds}. Here we will simply assume that
the critical region is sufficiently large to induce a significant focusing effect. We use the model of 
Nonaka and Asakawa \cite{Nonaka:2004pg} to describe the influence of the critical point on 
the thermodynamic variables. In this model the entropy density is obtained by interpolation between
the entropy densities $s_{\rm H},s_{\rm Q}$ of the hadronic and quark phase as
\begin{eqnarray}
s(T,\mu_B) &=& \frac{1}{2}\left(1-\tanh S_c\right) s_{\rm H}(T,\mu_B) 
\nonumber \\
&& + \frac{1}{2}\left(1+\tanh S_c\right) s_{\rm Q}(T,\mu_B) ,
\label{eq:entropy}
\end{eqnarray}
where $S_c(T,\mu_B)$ is proportional to the critical part of the entropy density obtained by rescaling the
expressions from the 3-dimensional Ising model. The proportionality constant entering into 
the definition of $S_c$ determines the size of the influence region of the critical point. Here 
we differ from ref.~\cite{Nonaka:2004pg} by choosing the parameters $\Delta T_{\rm crit} = 20\; {\rm MeV}, 
\Delta \mu_{\rm crit} = 100\; {\rm MeV}, D = 0.5$. This choice yields a much narrower critical region 
as shown in Fig.~\ref{fig1}. The width of the cross-over between the two phases at $\mu_B=0$ is 
approximately 45 MeV, in rough agreement with lattice-QCD results \cite{Aoki:2006br}.

\begin{figure}[htb]
\centerline{\includegraphics[width=0.95\linewidth]{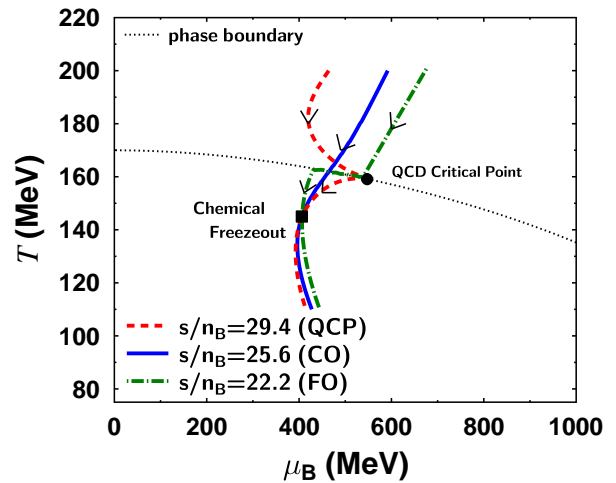}}
\caption{Hydrodynamical trajectories in the QCD phase diagram with and without the presence
of a critical point. Possible trajectories in the $\mu_B-T$ plane in the absence of a critical point are 
shown as solid line (for a cross-over transition) or dash-dotted line (for a first-order transition); 
the trajectory in the presence of a critical point is shown as dashed line. All trajectories meet at 
the bulk chemical freeze-out point. Arrows indicate the direction of time evolution.}
\label{fig2}
\end{figure}

The main characteristic of the fireball evolution in the presence of a critical point is that 
hydrodynamical trajectories, which would normally tilt to the right after crossing the phase boundary 
(see solid line in Fig.~\ref{fig2} for a smooth cross-over or the dash-dotted line for a first-order transition), 
make a detour into the vicinity of the critical point and then turn to the left as the temperature falls 
below $T_c$ (see dashed line in Fig.~\ref{fig2}).  For our argument, the important difference is the 
behavior just below $T_c$, where both $T$ and $\mu_B$ decrease for the trajectory through the 
critical point, while $\mu_B$ stays roughly constant or increases with falling temperature for 
trajectories away from the critical point. This difference can have visible consequences if hadron 
emission occurs over a finite range of temperatures, and if emission from different points along the 
trajectory can be discriminated. For instance, the ratio $\mu_B/T$ monotonically increases below 
$T_c$ along the ``normal'' (solid or dash-dotted) trajectories in Fig.~\ref{fig2}, implying a falling 
antiproton-to-proton ($\bar{p}/p$) ratio. On the other hand, the dashed trajectory in Fig.~\ref{fig2}
implies an approximately constant or even slightly decreasing value of $\mu_B/T$ and thus a
rising $\bar{p}/p$ ratio as the temperature falls below $T_c$. 

\begin{figure}[htb]
\centerline{\includegraphics[width=0.95\linewidth]{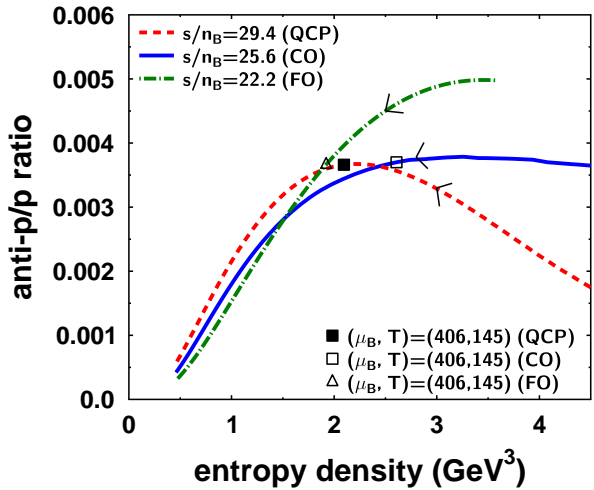}}
\caption{Antiproton-to-proton ratio along the trajectories shown in Fig.~\ref{fig2} as a function 
of the entropy density. The curves start at the phase boundary $T_c \approx 160$ MeV 
and continue down to $T \approx 110$ MeV. The location of the chemical freeze-out point 
$(\mu_{\rm ch},T_{\rm ch})$ deduced from experimental data is indicated by the open and solid 
squares. Note that the $\bar{p}/p$ ratio fonly rises for the trajectory deformed by the critical point.}
\label{fig3}
\end{figure}

In order to confirm this qualitative argument we present a quantitative analysis based on the 
assumption that the attractive basin of the critical point is reached in central Pb+Pb collisions 
at 40 GeV/$A$. In Fig.~\ref{fig3} we show the $\bar{p}/p$ ratio along the three trajectories 
shown in Fig.~\ref{fig2} as a function of the entropy density between $T_c$ and the chemical 
freeze-out point, which has been determined to lie at 
$(\mu_{\rm ch},T_{\rm ch}) \approx (400,145)$ MeV by a statistical model fit to experimental data 
\cite{BraunMunzinger:2003zd}. As anticipated, the $\bar{p}/p$ ratio falls or remains constant
between the phase boundary and chemical freeze-out for the ``normal'' trajectories (solid and 
dash-dotted lines), but rises for the trajectory deformed by the presence of the critical point 
(dashed line).

We next discuss how baryon emission from different points along the hydrodynamical trajectory
may be distinguished. We first note that data from Au+Au collisions at RHIC have been explained
by the assumption that the emission of hadrons with intermediate transverse momentum ($p_{\rm T}
\sim 2-5$ GeV/c) occurs at the phase boundary by recombination of constituent quarks 
\cite{Fries:2003vb,Greco:2003xt}. Bulk freeze-out of hadrons, on the other hand, occurs when the 
mean free path of hadrons becomes comparable to the size of the fireball. The mean free path 
relevant to transport properties generally grows with increasing hadron momentum. This implies 
that hadrons with large transverse momentum should freeze out earlier, on average, than hadrons 
with a small transverse momentum. In the extreme, intermediate $p_{\rm T}$ hadrons may be produced
at or near the phase boundary. This effect can also be understood by invoking detailed balance. 
A highly energetic hadron, impinging onto the fireball from the outside, would penetrate deeper 
into the matter than a low-energy hadron. Conversely, energetic hadrons will be emitted, on average 
from deeper inside the matter and thus earlier than low-energy hadrons. 

The differential emission of baryons as a function of transverse momentum can be analyzed 
quantitatively in the framework of a microscopic hadron transport model, e.~g.\ UrQMD 
\cite{Bass:1998ca,Bleicher:1999xi}. Such transport models based on relativistic Boltzmann dynamics
involving binary hadronic reactions are commonly used to describe the freeze-out and break-up of 
the fireball produced in relativistic heavy-ion collisions into hadrons. Utilizing the UrQMD model,
we have calculated central Au+Au collisions at a fixed target energy of 40 GeV/$A$, which may 
lead to conditions for which the matter passes near the QCD critical point. We then determined 
the  last time of interaction in the medium (``emission time'') for all final-state (anti-)protons. 
In order to discriminate between {\em slow} and {\em fast} hadrons, we choose two transverse
velocity windows:  $\beta_{\rm T} < 0.25$ for {\em slow} hadrons and $0.5 < \beta_{\rm T} < 0.75$
for {\em fast} hadrons, respectively.

\begin{figure}[htb]
\centerline{\includegraphics[width=0.95\linewidth]{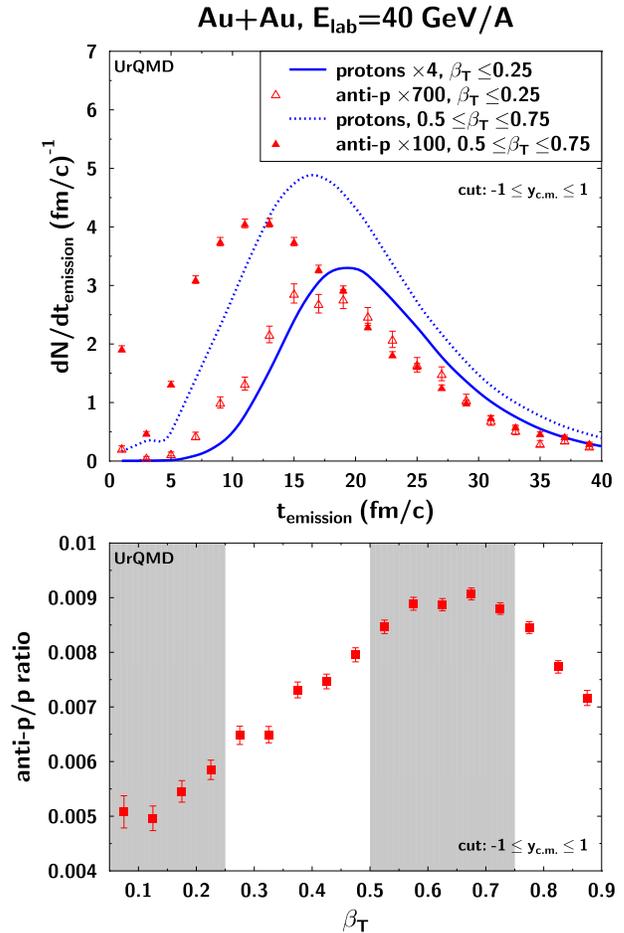}}
\caption{Top: UrQMD predictions for the emission time (last interaction time) distribution of protons 
and antiprotons in central Au+Au collisions at 40 GeV/$A$. The emission time distributions are 
shown separately for the kinematic windows $\beta_{\rm T} < 0.25$ and $0.5 < \beta_{\rm T} < 0.75$. Fast
particles are emitted on average significantly earlier than slow particles. 
Bottom: $\bar{p}/p$ ratio as a function of transverse velocity $\beta_{\rm T}$.}
\label{fig4}
\end{figure}

The distribution of emission times (counting from the moment of full overlap of the colliding 
nuclei) for {\em fast} (anti-)protons is shown in the top frame of 
Fig.~\ref{fig4} compared to the emission time distribution for {\em slow} (anti-)protons.  
The emission of protons and
antiprotons in the  $0.5 < \beta_{\rm T} < 0.75$  window is clearly correlated with 
early emission times and occurs 
approximately 4-5 fm/c earlier than for slow (anti-)protons. The average emission times for fast 
protons and antiprotons are $19.5 \pm 0.02$ and $16.5 \pm 0.06$ fm/c, respectively, compared 
to $22.9 \pm 0.01$ and $21.2\pm 0.2$ fm/c for slow moving protons and antiprotons. 

The bottom frame of Fig.~\ref{fig4} confirms the expectation from the solid line in Fig.~\ref{fig3},
that the $\bar{p}/p$ ratio should rise as a function of $\beta_{\rm T}$ in the absence of a critical point. Because the UrQMD calculation does not include finite-density corrections to the
$p+\bar{p}$ annihilation cross section, the overall value of the ratio should not be compared
with experiment. We also note that UrQMD does not contain any physics related to the QCD 
phase transition and only serves here as a model to study the correlation between emission 
time and transverse velocity. We expect the observed correlation to persist in the presence of 
a critical point. The fall in the $\bar{p}/p$ ratio for $\beta_{\rm T}>0.8$ is due to kinematical
constraints on the $\bar{p}$ production and independent of collision size and energy.
Since only a very small fraction of the baryon yield resides in the $\beta_{\rm T}>0.75$ range
we have selected the $0.5 < \beta_{\rm T} < 0.75$ range as representative for high $\beta_{\rm T}$
and early freeze-out nucleons.

A search of existing data revealed that the $\bar{p}/p$ ratio in Pb+Pb collisions at the CERN-SPS 
has been measured as a function of beam energy by the NA49 collaboration, which has published 
proton and antiproton spectra for fixed-target beam energies of 20, 30, 40, 80, and 158 GeV/$A$ 
\cite{Alt:2006dk}. Interestingly, the antiproton spectrum measured at 40 GeV/$A$ exhibits 
an anomaly. Whereas the exponential slope of the antiproton spectrum is flatter than the slope of 
the proton spectrum at other beam energies, it is slightly steeper at 40 GeV/$A$. 
A flatter antiproton spectrum is compatible with differential chemical freeze-out on a trajectory 
similar to the solid or dash-dotted trajectories shown in Fig.~\ref{fig2}, while a steeper spectrum 
would require a trajectory of the type expected in the vicinity of the critical point (similar
to the dashed line in Fig.~\ref{fig2}). The relative suppression of antiprotons at large transverse 
momentum is clearly visible in the spectrum itself (see Fig.~3 in ref.~\cite{Alt:2006dk}).
The size of the statistical errors of the measurement does not permit a firm conclusion about 
this anomaly, but it is certainly compatible with the arguments presented here.

In summary, we have shown that the evolution of the $\bar{p}/p$ ratio along isentropic curves
between the phase boundary in the QCD phase diagram and the chemical freeze-out point is
strongly dependent on the presence or absence of a critical point. When a critical point exists, 
the isentropic trajectory approximately corresponding to hydrodynamical expansion is deformed,
and the $\bar{p}/p$ ratio grows during the approach to chemical freeze-out. If nucleons of high
transverse momentum are chemically frozen out earlier than the slow nucleons, as it is suggested 
by microscopic simulations of hadronic dynamics, this result will translate into a $\bar{p}/p$ ratio
that falls with increasing transverse momentum instead of a rise or flat behavior in scenarios without 
critical point. This behavior would only occur at those beam energies, for which the fireball reaches 
the critical point. Depending on the actual size of the attractive region around the critical point, 
the search for an anomaly in the $y_{\rm T}$ dependence of the $\bar{p}/p$ ratio may require 
small beam energy steps. Note that the location of the critical point in our model study was chosen 
such that it is encountered by the hydrodynamical trajectory for conditions reached for a beam 
energy of 40 GeV/$A$ and a fixed-target. For a different location of the critical point, similar behavior 
would occur at other beam energies. 

{\it Acknowledgments:} 
This work was supported in part by grants from the U.~S.~Department of Energy, the Japanese 
Ministry of Education, and the Mitsubishi Foundation. 
We are grateful for the hospitality and support of the Yukawa Institute 
for Theoretical Physics in Kyoto during the workshop program entitled 
{\em New Frontiers in QCD 2008}. 
Finally, we thank G.~Odyniec for asking the question which inspired this investigation.


\begin{thebibliography}{10}

\bibitem{Aoki:2006we}
Y.~Aoki, G.~Endrodi, Z.~Fodor, S.~D. Katz, and K.~K. Szabo,
\newblock Nature {\bf 443}, 675 (2006).

\bibitem{Fodor:2002km}
Z.~Fodor, S.~D. Katz, and K.~K. Szabo,
\newblock Phys. Lett. {\bf B568}, 73 (2003).

\bibitem{Fodor:2004nz}
Z.~Fodor and S.~D. Katz,
\newblock JHEP {\bf 04}, 050 (2004).

\bibitem{Gavai:2004sd}
R.~V. Gavai and S.~Gupta,
\newblock Phys. Rev. {\bf D71}, 114014 (2005).

\bibitem{Stephans:2006tg}
G.~S.~F. Stephans,
\newblock J. Phys. {\bf G32}, S447 (2006).

\bibitem{Lacey:2006bc}
  R.~A.~Lacey {\it et al.},
  Phys.\ Rev.\ Lett.\  {\bf 98}, 092301 (2007).

\bibitem{Laszlo:2007ib}
A.~Laszlo {\it et al.}  [NA61 Collaboration],
arXiv:0709.1867 [nucl-ex].
  
\bibitem{deForcrand:2007rq}
P.~de~Forcrand, S.~Kim, and O.~Philipsen,
\newblock PoS {\bf LAT2007}, 178 (2007).

\bibitem{Stephanov:1998dy}
M.~A. Stephanov, K.~Rajagopal, and E.~V. Shuryak,
\newblock Phys. Rev. Lett. {\bf 81}, 4816 (1998).

\bibitem{Stephanov:1999zu}
M.~A. Stephanov, K.~Rajagopal, and E.~V. Shuryak,
\newblock Phys. Rev. {\bf D60}, 114028 (1999).

\bibitem{Berdnikov:1999ph}
B.~Berdnikov and K.~Rajagopal,
\newblock Phys. Rev. {\bf D61}, 105017 (2000).

\bibitem{Roland:2005pr}
NA49, C.~Roland,
\newblock J. Phys. {\bf G31}, S1075 (2005).

\bibitem{Roland:2006zz}
NA49, C.~Roland,
\newblock PoS {\bf CFRNC2006}, 012 (2006).

\bibitem{Nonaka:2004pg}
C.~Nonaka and M.~Asakawa,
\newblock Phys. Rev. {\bf C71}, 044904 (2005).

\bibitem{Hatta:2002sj}
Y.~Hatta and T.~Ikeda,
\newblock Phys. Rev. {\bf D67}, 014028 (2003).

\bibitem{Schaefer:2006ds}
B.-J. Schaefer and J.~Wambach,
\newblock Phys. Rev. {\bf D75}, 085015 (2007).

\bibitem{Aoki:2006br}
Y.~Aoki, Z.~Fodor, S.~D.~Katz and K.~K.~Szabo,
\newblock Phys.\ Lett.\  B {\bf 643}, 46 (2006).

\bibitem{Fries:2003vb}
R.~J.~Fries, B.~M\"uller, C.~Nonaka and S.~A.~Bass,
Phys.\ Rev.\ Lett.\  {\bf 90}, 202303 (2003).

\bibitem{Greco:2003xt}
V.~Greco, C.~M.~Ko and P.~Levai,
Phys.\ Rev.\ Lett.\  {\bf 90}, 202302 (2003).
  
\bibitem{Bass:1998ca}
S.~A.~Bass {\it et al.},
Prog.\ Part.\ Nucl.\ Phys.\  {\bf 41}, 255 (1998).

\bibitem{Bleicher:1999xi}
  M.~Bleicher {\it et al.},
  J.\ Phys.\ G {\bf 25}, 1859 (1999).

\bibitem{BraunMunzinger:2003zd}
P.~Braun-Munzinger, K.~Redlich, and J.~Stachel,
\newblock nucl-th/0304013.

\bibitem{Alt:2006dk}
  C.~Alt {\it et al.}  [NA49 Collaboration],
  Phys.\ Rev.\  C {\bf 73}, 044910 (2006).

\end{thebibliography}
\end{document}